\documentclass[pdflatex,sn-aps]{sn-jnl}
\usepackage{fix-cm}
\usepackage{titlesec}
\titlespacing*{\subsection}{0pt}{5pt}{5pt}
\usepackage{graphicx} 
\usepackage{multirow}%
\usepackage{amsmath,amssymb,amsfonts}%
\usepackage{amsthm}%
\usepackage{mathrsfs}%
\usepackage{xcolor}%
\usepackage{textcomp}%
\usepackage{manyfoot}%
\usepackage{booktabs}%
\usepackage{algorithm}%
\usepackage{algorithmicx}%
\usepackage{algpseudocode}%
\usepackage[utf8]{inputenc}
\usepackage{csquotes}
\usepackage[normalem]{ulem}
\usepackage{listings}\usepackage{rotating} 
  \usepackage{tcolorbox}
\usepackage[
singlelinecheck=false 
]{caption}
\hypersetup{
  pdftitle={Climate, Science, and the FCC: Think Clearly, Choose Wisely, Act Swiftly},
  pdfauthor={Patrick Janot and Christophe Grojean},
  pdfsubject={Climate change, fundamental research, and CERN's Future Circular Collider},
  pdfkeywords={FCC, CERN, climate change, fundamental research, particle physics, sustainability}
}

\usepackage{etoolbox}   
\makeatletter
\patchcmd{\@maketitle}{\artauthors}{\centering{\artauthors}}{}{}
\makeatother

\begin{document}
\title{Climate, Science, and the FCC}
\subtitle{\it Think Clearly, Choose Wisely, Act Swiftly}

\author*[1]{\fnm{Patrick} \sur{Janot}}\email{patrick.janot@cern.ch}
\author[2,3]{\fnm{Christophe} \sur{Grojean}}

\affil[1]{\small \orgname{CERN}, \orgdiv{EP Department}, \orgaddress{\street{1 Esplanade des Particules}, \city{Meyrin}, \country{Switzerland}}}

\affil[2]{\small \orgname{Deutsches Elektronen-Synchrotron DESY}, \orgaddress{\street{Notkestr. 85}, \postcode{22607} \city{Hamburg}, \country{Germany}}}

\affil[3]{\small \orgname{\orgdiv{Institut für Physik}, Humboldt-Universit\"at zu Berlin}, \orgaddress{\postcode{12489} \city{Berlin}, \country{Germany}}}

\abstract
{
Could a collider become the symbol of everything that must be abandoned to save the climate? In the summer of 2026, this question ran through discussions within the Particles and Fields Division of the French Physical Society. An unacceptable carbon footprint, two EPR nuclear reactors, “five hundred deaths”, billions to be reallocated and, to top it all, unclear scientific objectives: the objections to CERN’s Future Circular Collider (FCC) appeared to form an overwhelming indictment. This essay examines each charge in turn. Grounded in a clear and compelling scientific vision, it confronts emotionally charged imagery with orders of magnitude, distinguishes a project’s footprint from its full balance sheet, considers the costs of postponing or abandoning it, and explores ways of reducing its impacts, including the still uncertain prospect of natural hydrogen. The climate alone does not decide the future of the FCC; it requires us to judge the available choices by their real consequences. Developed for the public debate organised by France’s National Commission for Public Debate (CNDP) and for the Swiss consultation, this essay is intended for the broadest possible readership, well beyond particle physics and the scientific community. It frames the FCC debate as a choice about the kind of society we want to build: should we organise society around renunciation and ever-shrinking horizons, or transform how we produce and choose a future worth wanting?
}
\maketitle


\vfill
\noindent{\small\it \textbf{Foreword.}
The present text expands and unifies eight earlier essays~\cite{janot2026essays}; the collection includes their French originals, English translations and detailed supporting references. It contributes to both the CNDP public debate in France and the Swiss consultation on the FCC, while seeking to make the issues accessible to a wider audience. The views expressed, and any errors that remain, are the sole responsibility of the authors; they do not represent the position of CERN. The analysis is confined to climate change, which is only one dimension of the ecological crisis: water resources, soils and biodiversity require treatment in their own right.}
\eject

\section{The disaster, the trap, and the future we must build}\label{sec:disaster-trap-future}

In France, the seemingly endless heatwave of summer 2026 brought the climate emergency back into brutal focus. Forests were burning, the air was becoming unbreathable, and residents were being evacuated. How can we act swiftly and on the right scale to stop such disasters becoming the norm?

This question concerns political and economic decision-makers worldwide. It also concerns physicists. Like everyone else, they are living through these upheavals. But should they travel less and curtail their research? Should new scientific instruments, such as the Future Circular Collider envisaged at CERN, be abandoned?

The answer seems obvious. Since every tonne of greenhouse gases emitted contributes to warming, every tonne avoided would be a victory. And since a major project is more visible than a million scattered decisions, cancelling it would be exemplary, immediate, and beyond dispute.

But the obvious can be deceptive.

For decades, the global economy has extracted resources, manufactured and transported goods, all while emitting billions of tonnes of greenhouse gases. Then it presents each of us with the climate bill: here is the footprint~\cite{yoder2020footprint} of your home, your meals, your journeys, your purchases, and your holidays. Emissions are largely produced upstream; guilt is too often distributed downstream.

Faced with this status quo, a radical approach claims to offer a clean break~\cite{wiedmann2020affluence}. It turns that guilt into a programme: consume less, travel less, build less, communicate less, learn less—and ultimately research less; undertake nothing new; then single out a few symbols whose disappearance would prove that society had finally understood~\cite{scientifiques2025renoncer}. The individual no longer bears only the blame for the disaster. They must now shoulder almost alone the burden of solving it.

These two approaches present themselves as opposites, yet they share the same blind spot. The first allows systems of production to continue polluting while asking consumers to take responsibility for their footprint. The second instructs consumers, or institutions turned into scapegoats, to reduce their needs and ambitions until their footprint reaches a subjective threshold at which it would finally be acceptable. One assigns blame to everyone; the other imposes sacrifice upon them. Both focus on what is consumed, when what must be transformed is what is produced—and how.

\begin{quote}\itshape\normalsize
Rejecting this trap means neither denying the crisis nor claiming a right to limitless abundance.
\end{quote}
Some forms of waste must disappear, some practices must change, and every activity must reduce its impact. Sufficiency can help guide trade-offs. But it cannot become a doctrine of guilt, still less a blueprint for civilisation. The climate emergency 
requires us to transform the conditions under which eight billion human beings eat, receive healthcare, find shelter, work, learn, create, discover, share, play, live together, and hope.

A visible or exemplary action does not necessarily strike at the cause it claims to address. To fight climate change, we must compare orders of magnitude, identify the sources of emissions, and put symbols to the test of facts.

\begin{quote}\itshape\normalsize
The climate emergency commands us to act. Above all, it demands that we aim true. 
\end{quote}

\section{The FCC in the dock}\label{sec:fcc-in-the-dock}

A carbon footprint assigns emissions to a consumer, an activity, or an institution. When used to apportion blame, it diverts attention from the lever capable of eliminating those emissions. When electricity generated from coal powers a home, a hospital, a train, a factory, or a laboratory, each user inherits part of its footprint, yet none can replace the power station single-handedly. Confusing where an emission is accounted for with where it can be avoided identifies culprits without tackling causes.

The FCC offers fertile ground for this confusion. Because its footprint has been quantified and then made public~\cite{cern2026dmo,benedikt2025fsr3}, cancelling it may appear decisive. A strange paradox of transparency: what is measured becomes easier to condemn than less well-documented emissions. The project has a name, a timetable, a budget, and identifiable proponents. It can be isolated from the system on which it depends, cast as a guilty consumer, then made to answer alone for its projected footprint.

The case made against the FCC on the SFP forum and in a contribution to the public debate~\cite{scientifiques2026cndp} boils down to one question: Is the FCC compatible with the climate emergency? The main charges are these.

The first accusation is that its footprint is excessive: according to still-incomplete estimates, its construction and operation would emit between one and two million tonnes of carbon dioxide equivalent (CO$_2$e)~\cite{cern2026dmo,benedikt2025fsr3}. Its opponents argue that describing this figure as small in global terms would absolve the project of responsibility for its own emissions—a form of the rhetoric of inaction.

Its electricity consumption would also make it an energy ‘monster’, allegedly requiring the equivalent output of two EPR nuclear reactors, along with their attendant radioactive waste. Even the spectacular five-order-of-magnitude efficiency gain announced relative to LEP would, through a rebound effect, serve mainly to produce far more data while consuming as much energy—or even more.

Scientists, moreover, are said to have a duty to lead by example. Abandoning plans for a new research infrastructure would demonstrate that they are willing to subject themselves to the transformation demanded of society at large. The symbolic gesture would matter as much as, if not more than, the emissions avoided.

The project could also be deferred until more compact, less energy-intensive technologies become available. In the meantime, physicists could continue using existing facilities and analysing the data already collected. Making discoveries ten or twenty years later might seem a modest price to pay given the scale of the climate challenge.

Finally, the billions it would cost could fund research or action more directly useful to the low-carbon transition. Abandoning the FCC would therefore both avoid its emissions and free resources to tackle far greater ones.

Then comes a staggering accusation. According to a statistical rule associating one premature death with roughly 4,000 tonnes of CO$_2$e emitted~\cite{pearce2023deaths}, the FCC would cause ‘500 deaths’. Cancelling it would therefore no longer be merely a matter of climate policy. It would become a moral imperative.

If these criticisms were persuasive, it would also be because the accelerator’s scientific objectives appeared unclear to some. For decades, collider experiments were guided by the Standard Model, which offered something close to a guarantee of discovery. That endeavour culminated in the discovery of the Higgs boson, completing the set of particles it predicted. The LHC, however, is said to have revealed no clear signpost towards a deeper theory, while competing models now point in different directions. 

Nothing would then guarantee that the FCC would discover a new particle or confirm a theory beyond the Standard Model; it might merely measure already known particles with extreme precision. Would such a programme be enough to justify a new collider of this scale, cost and environmental impact? And before producing still more data, should we not first use increasingly powerful computing and artificial intelligence to reanalyse the immense quantities already archived, searching for signals or patterns that conventional methods may have missed?

All these objections converge on a seemingly unanswerable argument: since the FCC has a footprint, the surest way to eliminate that footprint would be to eliminate the project itself. A complex scientific, technological, and political decision is thus reduced to a simple subtraction.

But a footprint is not a balance sheet. It measures some adverse effects, yet says nothing about what an activity contributes, what would be lost if it were abandoned, or what would replace it. Reducing the climate impact of the FCC must remain an objective. Deciding its future also requires weighing this impact against the expected benefits, then comparing that balance with those of real alternatives. Refusing this second half of the calculation does not protect the climate: it writes the verdict before the case is heard.

The indictment has itemised the charges. We must now examine them, then draw up the balance sheet.
%
At times, this exercise offers a textbook illustration of Brandolini’s law: a one-sentence claim may require several pages of analysis before it can be properly assessed.

\section{The facts deliver their verdict}\label{sec:facts-verdict}

\subsection{Clear questions, open answers~\cite{benedikt2025fsr1}.} The absence of a guaranteed headline-making discovery does not make a scientific programme unclear. It means that the answer has not been written in advance.

The FCC’s purpose lies in a handful of fundamental questions. What is the world made of? Why does the Universe contain matter rather than antimatter? What is the nature of dark matter, which is five times more abundant than ordinary matter? Are the laws we know nature’s final word, or merely the first language in which we have learnt to describe it?

The discovery of the Higgs boson at CERN in 2012 revealed the last particle the Standard Model needed to describe known matter with remarkable precision. But this model does not tell the whole story. The questions it leaves open rank among the deepest in physics and cosmology. We know that our description is incomplete; we simply do not know how nature will reveal its limits. The LHC has shown us the way forward: both to scrutinise the world we know more closely and to explore beyond its frontiers.

Nor can existing knowledge, however comprehensively assembled, replace experiment. Artificial intelligence will become ever more powerful at connecting and synthesising existing knowledge. But it cannot determine by reasoning alone what nature does where no experiment has yet ventured. Nor can we. The rise of AI therefore makes experimental frontiers more important, not less. To cross them, we must continue to build scientific instruments at the necessary scale—and look.

Reanalysing existing data is, of course, invaluable. It is already being done within FCC teams using data collected at LEP, the FCC-ee’s predecessor at CERN, both to search for overlooked signals and to train new generations of physicists. But no method, however powerful, can extract from an archive events that the original experiment never produced. By recording, over a comparable period, event samples up to 100,000 times larger, and by operating at energies never before reached in electron–positron collisions, high enough to produce Higgs bosons and top quarks in abundance, the FCC-ee would bring the known world into focus with unprecedented clarity. 

With data on this scale, it might discover new particles, some light but so elusive that they have escaped us until now, and others too massive to reveal themselves directly but detectable through the subtle traces they would leave in precision measurements, or uncover fundamental laws still unknown to us. Each offers a possible path towards shedding light on questions that remain unresolved. A proton collider would then take up the search at far higher energies.

Yet if no deviation appeared after analysing trillions of decays, entire families of explanations would be ruled out. Nature would be telling us that its next layer of organisation lies further away, is more elusive, or is more radically different than we imagine.

\begin{quote}\itshape\normalsize
Measured with sufficient precision, even nature’s silence is an answer.
\end{quote}

The FCC cannot therefore simply ‘find nothing’. It will either reveal a path towards new physics or show that many of the paths we had imagined are dead ends, clearing the way towards a fuller understanding of nature. In either case, our knowledge will have advanced.

\begin{quote}\itshape\normalsize
The questions are clear. What remains open, as it must in any genuine exploration, is nature’s answer.
\end{quote}

\subsection{Changing scale~\cite{janot2026orders,janot2026verdict}.} 

The carbon footprint attributed to the construction and operation of the FCC-ee amounts to roughly 40,000 tonnes of CO$_2$e per year over 35 to 40 years~\cite{cern2026dmo,benedikt2025fsr3}. On its own, the figure is striking. Divided by the world’s eight billion people, whom this scientific infrastructure is intended to benefit, it comes to a few grams per person per year.\footnote{The FCC brings together scientists from around the world; the knowledge and innovations it will produce will circulate far beyond CERN. Relating its footprint to the world's population thus provides a common denominator for gauging its order of magnitude against global climate challenges. This calculation implies neither that consumption or benefits are uniformly distributed nor that each person bears individual responsibility for the project's emissions.}

\begin{quote}\itshape\normalsize
A few grams are not zero. But they do not spell climate catastrophe.
\end{quote}

To understand how the aggregation effect arises, we must look at global emissions. In 2024, human activities emitted nearly 58 billion tonnes of CO$_2$e. Electricity generation alone accounted for almost 16 billion tonnes, more than a quarter of the total~\cite{unep2025emissionsgap}. Roughly 15 of those 16 billion tonnes, i.e. more than 95\%, still came from coal, gas, oil, and other carbon-intensive sources~\cite{ember2025explorer}. By contrast, nuclear power, hydropower, wind, and solar already supplied nearly 40\% of the world’s electricity~\cite{ember2025review}, with associated emissions of approximately 250 million tonnes of CO$_2$e per year~\cite{authors2026estimate}.

Electricity generation therefore accounts for nearly two tonnes of CO$_2$e per person per year, with about 1.9 tonnes coming from carbon-intensive sources.

\begin{quote}\itshape\normalsize
A few grams for the FCC; two tonnes for global electricity generation.
\end{quote}

Generating as much electricity while progressively replacing fossil-fuel sources with low-carbon alternatives would cut the associated emissions by about 95\%, avoiding more than 12 billion tonnes a year. Cancelling the FCC would avoid about 40,000 tonnes a year. Decarbonising global electricity generation would therefore avoid roughly 300,000 times as much CO$_2$e as sacrificing the project.

\begin{quote}\itshape\normalsize
Three hundred thousand. In climate action, scale outweighs symbolism.
\end{quote}

This comparison is not a licence to pollute. If every sector split its emissions so that each part could be declared negligible, no one would ever act. Every sector must reduce its emissions. But a climate strategy must rank the levers available to it: every reduction counts, yet not all have the same reach. Prioritising does not mean absolving.

Above all, we must compare two courses of action. Cancelling the FCC would remove one particular use of electricity. Decarbonising electricity generation would simultaneously reduce the footprint of trains, homes, hospitals, factories, data centres, household appliances, and scientific facilities connected to the grid. The first removes a use. The second transforms the consequences of billions of uses—not by ceasing to produce the electricity on which our societies depend, nor by asking everyone to reduce their consumption by a factor of ten, but by replacing the most polluting forms of generation.

Replacing an incandescent bulb with an LED in your kitchen is sensible. Claiming thereby to have retrofitted an entire city would be an illusion: the insulation of its buildings, their heating systems, and the generation of the electricity supplying them would remain unchanged. Likewise, reducing the FCC footprint is essential; presenting its cancellation as a climate strategy would repeat the same mistake.

Numbers do not decide what deserves to exist. They prevent us from mistaking a largely symbolic gesture for a response commensurate with the challenge.

\begin{quote}\itshape\normalsize
When it comes to climate, getting the scale wrong means getting the priority wrong.
\end{quote}

\subsection{From example to catalyst.} 

For science, setting an example does not mean disappearing, but leading. It begins by pooling equipment, expertise, and resources in a single infrastructure used by a global community, rather than dispersing them.

\begin{quote}\itshape\normalsize
Science does not owe society a penance. It owes it a contribution.
\end{quote}

Like any infrastructure, the FCC will be a beneficiary of the low-carbon transition. As electricity, steel, cement, and transport decarbonise, its footprint will shrink. It must accelerate this trend by favouring the lowest-emission processes and reducing its own resource needs.

But above all, it can become a catalyst. The performance demands of accelerators drive advances in energy efficiency, cryogenics, and high-temperature superconductivity. Preparations for the FCC are already stimulating research into low-carbon cements, more resource-efficient civil-engineering methods, and the reuse of excavated materials~\cite{hauzinger2025excavation,voiron2022molasse}. CERN has neither shareholders to reward nor technological rents to preserve: its purpose is not profit but knowledge. The knowledge gained and technologies developed in pursuit of its research are intended to spread and benefit society.

These catalytic effects will need to be demonstrated. Measurable objectives must be defined, knowledge transfer must be organised, and outcomes must be evaluated. When an innovation developed for a scientific infrastructure is deployed in thousands of factories or millions of devices, its impact may far exceed the footprint of the project that enabled it. 

That is the example worth setting: improving the wider balance sheet. From the preparatory stage onwards, the FCC is committed to action on both fronts: reducing its future footprint and helping to transform the systems on which all footprints depend. That commitment must be put into practice throughout construction and operation.

\begin{quote}\itshape\normalsize
The climate crisis will not be overcome by sacrificing symbols. It compels us to transform our means of action.
\end{quote}

\subsection{Into the valley of Death rode the ‘five hundred’~\cite{janot2026hope}.} The ‘500 deaths’ cited in the indictment come from applying a statistical rule of one premature death for every 4,000 tonnes of CO$_2$e emitted to the emissions attributed to the FCC over 40 years. The figure carries the force of a verdict. But what exactly does it measure?

This rule does not identify anyone whose death would be directly caused by the FCC. Based on a chain of assumptions and uncertainties, it statistically distributes a fraction of global climate risk over several decades and across vast populations. Its authors acknowledge that the result could be multiplied or divided by ten~\cite{pearce2023deaths}. Converting emissions into human lives lost therefore makes their consequences neither more certain nor better understood. It merely makes them harder to discuss without appearing insensitive. The emotional charge of such wording can demoralise and paralyse its audience, at the risk of fostering the very inaction it claims to oppose.

This metric can help compare global emissions scenarios and serves as a reminder that every emission contributes to climate risk. It becomes misleading when used, in isolation, to determine which human activities should disappear. Applied mechanically, it would also attribute deaths to a hospital, a university, or a solar-panel factory, without counting the lives those infrastructures might save. A serious decision requires a full balance sheet.

Here the fundamental asymmetry becomes apparent. Some of the FCC’s adverse impacts can be estimated before it is built. Some of its future benefits can be anticipated, but not quantified with equal confidence. Yet technologies developed for accelerators and detectors have already contributed to medical imaging, nuclear medicine, and cancer treatment. We must also count the knowledge produced, the training provided through research, the peaceful collaboration of scientists from around the world united by shared values, and innovations capable of reducing emissions far beyond CERN.

It would be artificial to credit the FCC in advance with all the lives that such advances might help to save~\cite{janot2026hope}. It is just as artificial to charge it immediately with the calculated death toll, then set at zero everything that the same accounting system cannot measure on the assets side.

\begin{quote}\itshape\normalsize
Quantifying uncertain harms does not make them certain. Not yet being able to quantify future benefits precisely does not make them non-existent.
\end{quote}

The count of premature deaths should be neither mocked nor dismissed. It must be restored to its proper status: not a list of identified victims of the FCC, but an uncertain rendering of a statistical fraction of global climate risk. A figure that arouses emotion must be handled with the greatest care: when it takes the place of reasoning, it becomes an instrument of manipulation.

\begin{quote}\itshape\normalsize
A metric, however striking, never tells the whole story on its own: counting deaths is not enough to decide how to save lives.
\end{quote}

\subsection{Is time free?~\cite{grojean2026time}} 

More efficient, less resource-intensive future technologies might reduce the footprint of a postponed FCC. But waiting for them means trading a mature scientific programme for the risky wager that the same knowledge could one day be produced at a lower environmental cost.

The FCC is more than a machine. It is a scientific community, technical expertise, training programmes, rare skills, and decades of international cooperation. Postponing it does more than shift a date in the calendar: teams disperse, expertise is lost, young talent takes other paths, and collective momentum erodes. A society cannot prepare for the long term and respond to crises if it treats science as a variable to be adjusted according to the pressures of the present.

After twenty years of European planning for the post-LHC era, conditions rarely found together have now converged: broad consensus within the particle physics community, international mobilisation around a shared scientific vision~\cite{cerncouncil2026strategy}, structured dialogue with the authorities of the Host States, and concrete funding prospects, including from the European Union~\cite{ec2025horizoneurope} and private donors~\cite{cern2025donors}. This alignment cannot be taken for granted: there is no guarantee that it could be recreated ten or twenty years from now.

For a project whose timescale is already measured in decades, such a delay would not be a mere interlude: it would cost science a generation. Time is not a waiting room where knowledge, teams, and opportunities sit patiently.

\begin{quote}\itshape\normalsize
To postpone is too often to engineer renunciation without ever naming it.
\end{quote}
\subsection{The mirage of freed-up resources~\cite{janot2026mirage}.} Could the billions intended for the FCC be redirected towards the energy transition or climate research? It is a reasonable hypothesis; past experience~\cite{janot2026mirage}, however, counsels against taking it for granted.

Founded after the Second World War as a project of peaceful scientific cooperation, CERN rests on long-term commitments by its Member States. Its mission of bringing countries together despite their differences remains essential today. Scaling back those commitments would be a political choice far broader than abandoning the FCC alone. For more than seventy years, the continuity of the Organisation’s recurrent funding has enabled the development of an ecosystem of laboratories, expertise, and international scientific cooperation.

The FCC would rely first and foremost on that recurrent funding, potentially supplemented by prospective European support for the project and private donations pledged for that purpose. Abandoning the FCC would neither turn CERN’s recurrent funding into a climate fund nor make project-specific funding freely available for reallocation. It could, however, weaken this painstakingly built ecosystem, while any resources it freed would not automatically transform into climate research laboratories. Their destination would depend on political choices whose outcome cannot be prejudged.


\begin{quote}\itshape
\normalsize
A euro taken from the FCC does not automatically strengthen climate action. It does, however, cease to support fundamental research and the scientific capabilities it creates.
\end{quote}

Treating research as a closed system forces a zero-sum game: climate versus particle physics, medicine versus accelerators, immediate needs versus fundamental knowledge, when the challenges of this century will require all these fields. Collective ambition attracts talent, collaborations, investment, and new sources of funding. Making one community’s disappearance a condition for another’s future legitimises a systemic weakening of research. The answer lies in increasing the resources available to research as a whole.

A partial reallocation would still be possible, but its climate impact would depend on what the freed-up resources actually funded. Abandoning the FCC would mean losing a scientific programme and the innovations it would stimulate. Its teams would not be reassigned by decree to research on the energy transition. Crediting cancellation with all avoided emissions and every hoped-for reallocation, while ignoring forgone benefits, is not drawing up a balance sheet. It is writing the conclusion in advance.

There is no scientific savings account holding money that might otherwise have funded the FCC. Budgets are approved each year; major scientific projects are built over decades. 

\begin{quote}\itshape\normalsize
Confusing these two timescales is politically naive: the hoped-for reallocations remain hypothetical; the losses would be all too real.
\end{quote}
\subsection{From the monster to the phantom EPRs~\cite{janot2026monster}.} Reducing the FCC’s electricity consumption, decarbonising its supply, and limiting the associated radioactive waste must remain priorities. But we must begin with the numbers, not with the image of a ‘monster’. Depending on the mode of operation, the FCC-ee would consume between 1.1 and 1.8 terawatt-hours (TWh)~\cite{benedikt2025fsr2}, averaging about 1.3 TWh a year—a considerable requirement for a single facility, but close to CERN’s current consumption when the LHC is operating. 

This is a substantial amount of electricity. For comparison, Meta’s data centre in Sarpy, Nebraska, consumed 1.258 TWh in 2024~\cite{meta2025data}. In the same year, 137 new hyperscale data centres of various sizes came online worldwide—more than one every three days~\cite{synergy2025hyperscale}. These comparisons do not absolve the FCC, but they show that its demand would not be unprecedented among large electricity-consuming infrastructures.

Nor would it enter an electricity-starved system: despite exceptional heatwaves and persistent drought, France remained a net electricity exporter for 99.9\% of the summer of 2026~\cite{rte2026summer}. Should the FCC nevertheless reduce its consumption? Certainly: no major consumer is exempt from the pursuit of efficiency. But the relevant question is not whether 1.3 TWh is a large number—it is. It is how much scientific value can be produced with that electricity, how it is generated, and when it is consumed.

An EPR is designed to produce 13 TWh a year~\cite{cde2024epr}. The arithmetic fits on one line: 1.3 for the FCC; 26 for two EPRs. The monster laboured and brought forth a mouse: the FCC would require only 5\% of their output. Even this 5\% overestimates the share associated with nuclear waste. CERN buys electricity from the French grid, generated mainly from low-carbon sources: hydropower, wind, solar, and nuclear power. If nuclear power supplied half of that mix in 2050, the waste-associated share would amount to only 2.5\% of the output of the two EPRs in the original claim.

Moreover, in 2024 CERN signed three long-term power purchase agreements, thereby enabling new solar facilities to be financed that will supply 140 GWh a year from 2027~\cite{cern2024solar}. The FCC will have to build on this strategy and place efficiency at the heart of its design.

The rebound effect cited in the indictment must be taken seriously. The FCC’s efficiency gains are intended primarily to achieve its scientific goals, not to reproduce those of LEP while consuming 100,000 times less energy. They do not, however, remove the need to control total consumption.

Finally, the impact of electricity consumption also depends on when it occurs. In a grid increasingly supplied by solar and wind power, the question will no longer be simply {\it How much electricity does the FCC consume?} but also {\it When does it consume it?}

Unlike a hospital, a collider need not operate continuously. CERN already adjusts its schedule to the grid’s seasonal constraints. The FCC could go further: operate flat out when renewable electricity is abundant, reduce its power draw at times of peak demand, and reschedule some operations when generation is scarce. A large, inflexible consumer increases strain on the grid; a flexible one can absorb surpluses and scale back when electricity demand is high. 

\begin{quote}\itshape\normalsize
The FCC must not merely consume less. It must consume at the right time.
\end{quote}

The indictment invoked two EPRs; the calculation brings the FCC’s requirement down to 5\% of their output. Assuming a future electricity mix of up to 80\% renewables, flexible operation prioritising periods of abundance could leave nuclear power supplying only 20–40\% of the FCC’s electricity—roughly 1–2\% of the output of the two reactors invoked. Flexible operation can reduce both pressure on the grid and the FCC’s residual nuclear share. It cannot, however, bridge every prolonged period of low wind and solar generation.

\begin{quote}\itshape\normalsize
Monsters feed fantasies. Orders of magnitude guide decisions.
\end{quote}

\subsection{Hydrogen mon amour~\cite{grojean2026hydrogen}.} Even at this reduced level, reliance on nuclear power would leave the question of radioactive waste unresolved. This is not a climate issue: nuclear power is among the lowest-carbon sources of electricity. But can this fraction be reduced further? Let us follow electricity from the simplest option to the most uncertain: consume it directly when available; draw, where it exists, on the grid’s hydropower flexibility; then shift surpluses by a few hours using sodium-ion~\cite{iea2026sodium,iea2024batteries} or second-life~\cite{bobba2018secondlife} batteries, whose round-trip efficiency can exceed 80\%~\cite{doe2023stateflow}. Yet prolonged periods with little wind or sun, the {\it dark doldrums}, require other reserves.

\begin{quote}\itshape\normalsize
Batteries shift electricity across hours; could hydrogen carry it across days?
\end{quote}

Green hydrogen, produced by electrolysis of water, can be stored and later converted back into electricity using a fuel cell or turbine. But each conversion takes its toll in energy and adds to its footprint: of every 100 kWh fed into the chain, fewer than 30 would return to the grid~\cite{iea2025ukrainehydrogen}. Using it every day to shift solar electricity from noon to evening would be absurd. Yet it could prove useful as a reserve, seldom called upon but available during {\it the dark doldrums}.

A long-overlooked possibility then changes the picture: hydrogen also occurs naturally underground. It need no longer be manufactured at the cost of successive conversions; it becomes a resource to extract, purify, and convert. Geological reactions may even continue to produce it, though we do not know at what rate or whether they could replenish a reservoir as it is exploited~\cite{ellis2024hydrogen}.

In Lorraine, hydrogen has recently been measured in a former coalfield, with concentrations increasing with depth. Some extrapolations suggest as much as 46 million tonnes~\cite{bettayeb2023hydrogen}. As an upper bound, replacing the FCC’s entire residual nuclear contribution over fifteen years would require about 1\% of that quantity, or roughly 45–90 tonnes a day. Sustained extraction at that scale has yet to be demonstrated from any natural hydrogen deposit. The use envisaged here would be far more limited: bridging the dark doldrums when renewables and batteries would fall short.

Its climate footprint also remains to be established: a natural resource is not automatically sustainable simply because it lies underground. Associated gases, leaks, purification, and drilling could greatly increase that footprint; extraction could also affect ecosystems and groundwater resources. 
%
Natural hydrogen will have to earn its low-carbon status with hard numbers.

If white hydrogen proved its potential, CERN could help establish a genuinely low-carbon industry without itself becoming a gas producer. A long-term purchase agreement modelled on the 2024 solar contracts would provide a producer with a guaranteed market. Extracted and converted near the deposit, the hydrogen would feed electricity into the grid when renewables and batteries fell short. The contract would impose verifiable criteria for production rate, leakage, cost, and carbon intensity.

\begin{quote}\itshape\normalsize
CERN would become a launch customer.
\end{quote}

Should the resource prove scarce, generating electricity would probably not be its best use. It should first replace fossil-derived hydrogen in fertiliser production, steelmaking, or the chemical industry, and serve hard-to-electrify sectors. Yet a tiny fraction used as a long-duration reserve could secure the FCC’s electricity supply even through the dark doldrums and further reduce the residual nuclear share—and with it, the associated radioactive waste.

The paradox would be fruitful: the FCC’s residual need could help create an industry capable of reducing emissions across entire sectors. Natural hydrogen is not yet the ‘white gold’ of the twenty-first century. To earn that title, it will have to flow sustainably from the ground at a measurable rate—not merely in press releases.

\begin{quote}\itshape\normalsize
In love as in science, promises are not enough. We need evidence.
\end{quote}

\section{Mobilising to match the threat~\cite{janot2026war}}\label{sec:mobilising}

Examining the facts helps us prioritise the levers for action and explore solutions. It does not decide for us. But time is running short: every year of inaction adds to a climate legacy whose effects will persist for decades. This urgency compels us to mobilise.

Every mobilisation begins by naming the threat. The threat lies neither in human needs, nor in people’s aspiration to escape poverty, nor in science or progress. It lies in the accumulation of greenhouse gases and in the processes that drive it.

This mobilisation cannot rest on blaming individuals or exemplary sacrifices, still less on waiting for a technological miracle. It calls for large-scale collective action across electricity generation, buildings, transport, steel, cement, the chemical industry, agriculture, and land use.

The first task is scientific and technological. We must develop and deploy low-carbon power technologies, energy storage, smart grids, resource-efficient materials, clean industrial processes, and the capture of residual emissions. Science does not guarantee a solution to every difficulty. Weakening research, however, guarantees that fewer solutions will be explored. Ignorance can be an instrument of power; it has never made a physical constraint yield.

But a technology confined to a laboratory cannot alter the climate trajectory. The second task is therefore industrial and financial. We must build infrastructure, train workers, secure supplies, and invest beyond electoral cycles. Public funding must reduce initial risks; standards and long-term contracts must create markets; industry must turn prototypes into accessible solutions.

There remains the political and democratic task. This transformation can succeed only through collective choices. Who should finance the transition? What must be transformed—and what do we want to preserve? What efforts should be required, of whom, and under what conditions? No equation can answer these questions. A material constraint calls for strong institutions, widely understood rules, fair burden-sharing, and a debate that distinguishes facts from preferences—not a moral lecture. Only then can climate action emerge with the strength to weather crises, survive changes of government, and stand up to vested interests.

The 2015 Paris Agreement~\cite{un2015paris} gave this shared responsibility global scope. It did not ask eight billion individuals to become climate heroes; it asked states to organise and coordinate their efforts. Its implementation remains far from adequate; its guiding principle remains sound: a global problem calls for collective commitments that can be verified and progressively strengthened.

In this shared effort, everyone has a responsibility—but no one should carry the weight of the world alone. An individual can avoid waste or change a habit. They are also a worker, a neighbour, a member of an association, a voter, and a citizen. Their power extends beyond their shopping basket and travel choices: they can change practices at work, support collective investment, take part in local decision-making, and choose those who will set the rules.

Physicists have a particular role here: to establish orders of magnitude, measure impacts, improve the efficiency of machines, work on energy, materials, and storage, and then inform public decision-making. Learned societies can pool expertise, publish accessible analyses, and distinguish solutions from promises. They can also help innovations move between laboratories, infrastructures, and industry, while fostering dialogue with other disciplines.

It is not for scientists alone to decide what society should build or abandon. Their expertise is no substitute for democratic decision-making; it must ensure that such decisions do not rest on erroneous figures, misleading comparisons, or physical impossibilities. The voice of science does not dictate. Its rigour belongs in public debate.

This mobilisation requires more civic engagement, more democracy, more science, and more international cooperation: no border can stop a molecule of CO$_2$. 

\begin{quote}\itshape\normalsize
Faced with a major crisis, a society does not scatter its forces. It marshals them.
\end{quote}

\section{Neither submit nor renounce: A future worth wanting}\label{sec:future-worth-wanting}

At the beginning of this essay, forests were burning. Wild animals were fleeing the flames while they still could. The air had become difficult to breathe. Entire families were leaving their homes without knowing what they would find on their return. One question was inescapable: where must we act now to prevent these recurring disasters from worsening and becoming our new normal?

The answer cannot be to carry on as before: the global system of production must stop releasing tens of billions of tonnes of greenhouse gases into the atmosphere each year. But nor can it be to present individuals with the bill, then offer them no future but a life defined by everything they must give up.

Between these two dead ends, a more difficult, and infinitely more ambitious, path remains open: transforming how we generate electricity, produce materials and goods, construct buildings, and travel. We will also need to reduce waste, make trade-offs between competing uses, and accept physical limits.

\begin{quote}\itshape\normalsize
Not by giving up on meeting the needs of human societies, but by learning collectively to meet them differently—and sustainably.
\end{quote}

This path must allow us to continue feeding ourselves, receiving healthcare, housing ourselves, working, learning, creating, understanding, and enjoying life—without destroying the very conditions of our prosperity or our ability to live in peace.

That is what a future worth wanting means. It promises neither endless material abundance nor the abolition of every constraint. It would recognise physical limits and share their burdens fairly, while preserving the ambition to devise ways of living together without overstepping them.


In that future, no one is reduced to their footprint. But freeing individuals from guilt does not mean issuing them with a certificate of innocence. Their responsibility does not disappear into collective action. It is there that it finally finds its power.

The FCC is just one example, but it puts this collective choice to the test. It must be neither shielded from scrutiny nor sacrificed. It must reduce its footprint, adapt to the electricity grid, support new renewable generation, advance the most efficient technologies, and help new solutions emerge. It must be judged on its full balance sheet: what it consumes, what it produces, what it transforms, and what, without it, would never see the light of day.


A civilisation facing an existential threat does not protect its future by diminishing its capacity to understand or by setting fields of knowledge against one another. It mobilises knowledge through cooperation and collective intelligence, giving every field of science the means to address present needs and help build the future we share.

Science alone will not save the climate. Technology cannot replace political courage, justice, or democracy. But without knowledge, politics goes blind; without facts, democracy becomes vulnerable. A society that renounces discovery and invention condemns itself to mere endurance.

Those who will live with the climate we leave them will not ask us whether we felt guilty enough. They will not count the symbols sacrificed to prove our virtue. They will ask whether we faced the facts, chose the levers powerful enough to change our course and, before it was too late, transformed the systems on which their lives will depend.

\begin{quote}\itshape\normalsize
They will ask whether, while the forests were burning, we organised our surrender—or built the future.
\end{quote}

The hour is late. But it is not too late to act, invent, cooperate, and choose. The laws of physics set the limits. They do not choose our future.

\begin{quote}\itshape\normalsize
That responsibility remains entirely our own.
\end{quote}

\vfill\eject
\section*{Acknowledgements}

We begin by acknowledging those whose arguments shaped the case against the FCC. They may not recognise the conclusions reached here, but without the clarity and, at times, the admirable vigour with which they pressed their case, this essay would have lacked much of its purpose. They gave us much to examine, carefully unpick or, where necessary, debunk, and ultimately transform into the starting point for a more constructive vision. Disagreement is not always this productive.

We are deeply grateful to Claire Adam, Nima Arkani-Hamed, Olivier Arnaez, Ursula Bassler, Laurent Baulieu, Nicolas Bellegarde, Olga Beltramello, Michael Benedikt, Gregorio Bernardi, Philippe Bloch, Alain Blondel, Frederick Bordry, Gaëlle Boudoul, Vincent Breton, Jean-Paul Burnet, Matteo Cacciari, Matthew Chalmers, Maurice Chapellier, Claude Charlot, Laurent Chevalier, Johann Collot, Alain Cordier, Pierre Darriulat, Michel Davier, Jacques Delabrouille, Marco Delmastro, David d'Enterria, Stéphanie Escoffier, Louis Fayard, Daniel Froidevaux, Fabiola Gianotti, Jean-Fran\c{c}ois Grivaz, Johannes Gutleber, Gautier Hamel de Monchenault, Sophie Henrot-Versillé, Michel Jouvin, Marumi Kado, Mattis Kennouche, Philippe Lebrun, Julien Le Gal Thomas, Jessica Lévêque, Sir Chris Llewellyn Smith, Carlos Louren\c{c}o, Michelangelo Mangano, Florian Nortier, Jacques Pichoff, Eliezer Rabinovici, Pasquale Serpico, Marie-Hélène Schune, Géraldine Servant, Yves Sirois, Michel Spiro, Vincent Tisserand, Daniel Treille, Laurent Vacavant and Frank Zimmermann. We also thank the many others who took the time to offer us words of encouragement in conversation.

In different ways, they encouraged us, read successive drafts with care, checked figures, challenged formulations, offered corrections and objections, or drew our attention to questions we had overlooked.  Their support sustained us, and their scrutiny made this essay more accurate, more balanced and, we hope, more persuasive.

We owe a special debt to Jacques Lefrançois, whose counsel helped shape the political thinking that runs through this essay.

\subsection*{Disclaimer} The French version of this essay was originally prepared by the authors for a French-speaking readership. Artificial-intelligence language tools assisted both in proofreading the French text and in translating it into English. The translation was reviewed by several native speakers of English. The authors remain fully responsible for the final text.

\bibliography{Tribune_CNDP}

\end{document}